\documentclass[prl,twocolumn,showpacs,preprintnumbers,amsmath,amssymb]{revtex4}

\usepackage{graphicx}
\usepackage{dcolumn}
\usepackage{bm}

\begin{document}

\preprint{{\em Supercond. Sci. Technol.} (2003), in press}

\title[To appear in {\em Supercond. Sci. Technol.} (2003)]{Properties of superconducting MgB$_2$ wires: ``in-situ" versus ``ex-situ" reaction technique}
\author{Alexey V. Pan, Sihai Zhou, Huakun Liu and Shixue Dou}
\affiliation{Institute for Superconducting and Electronic Materials, University of Wollongong, \\ Northfields Avenue, Wollongong, NSW 2522, Australia}
\date{\today}

\begin{abstract}

We have fabricated a series of iron-sheathed superconducting wires prepared by the powder-in-tube technique from (MgB$_2$)$_{1-x}$:(Mg+2B)$_x$ initial powder mixtures taken with different proportions, so that $x$ varies from 0 to 1. It turned out that ``ex-situ" prepared wire ($x = 0$) has considerable disadvantages compared to all the other wires in which ``in-situ" assisted ($0 < x < 1$) or pure ``in-situ" ($x = 1$) preparation was used due to weaker inter-grain connectivity. As a result, higher critical current densities $J_c$ were measured over the entire range of applied magnetic fields $B_a$ for all the samples with $x > 0$. Pinning of vortices in MgB$_2$ wires is shown to be due to grain boundaries. $J_c(B_a)$ behavior is governed by an interplay between the transparency of grain boundaries and the amount of ``pinning" grain boundaries. Differences between thermo-magnetic flux-jump instabilities in the samples and a possible threat to practical applications are also discussed.
\end{abstract}




\maketitle

\section{Introduction}
A simple composition, rather easy preparation, a relatively high critical temperature $T_c$, quite high critical currents, and low sensitivity to weak links make MgB$_2$ superconductor a suitable candidate for practical large-scale applications, such as long wires for power lines, magnets, motors, generators, etc. The powder-in-tube (PIT) technique has been adopted for wire fabrication with this brittle material. One approach to the fabrication is to use pre-reacted MgB$_2$ powder, the so called ``ex-situ" reaction preparation. High values of transport critical current $I_c$ in the ex-situ prepared tapes have been achieved even without further sintering \cite{grasso}. However, 1\% of additional impurity in the initial pre-reacted MgB$_2$ powder has been shown to completely suppress $J_c$ in tapes \cite{fujii}. It was further shown that a heat treatment of such tapes, following the mechanical deformation stage, can increase the critical current density $J_c$, as a consequence of stronger grain connectivity due to the heat treatment \cite{suo}. In this work \cite{suo}, only hard sheath materials were used, which are beneficial for the ex-situ preparation due to chemical compatibility and the possibility of achieving high core densities inside the sheath. Nevertheless, the poor ability of MgB$_2$ particles to form a bulk monolithic structure remains an important issue for further $J_c$ enhancement, as was shown, for example, by the two-stage wire fabrication process discussed in Refs.~\cite{pan2,pan3}. This process involved an initial in-situ wire preparation from an unreacted Mg+2B powder mixture, as the first stage, and a {\it quasi ex-situ} second stage consisting of an additional mechanical deformation of the wires, which crashed the core that had been formed in order to increase its density, followed by another heat treatment. The results indicated that wires made by the ex-situ preparation procedure are not optimal.

An alternative ``in-situ" reaction approach has been successfully used for achieving excellent superconducting characteristics for MgB$_2$ wires \cite{xiaolin,martinez}, in particular if SiC nano-particle doping is introduced \cite{pansic1,pansic2}. The in-situ preparation has, however, its own disadvantage: the density of the MgB$_2$ core after its formation is rather low, and its microstructure is extremely porous. This is a consequence of the phase transformation from Mg+2B $\rightarrow$ MgB$_2$, because the theoretical mass density of the initial Mg+2B mixture is significantly lower than that of the MgB$_2$ phase \cite{pan2}. As already mentioned, an additional stage of mechanic deformation and consecutive heat treatment did not help to overcome this problem.

In this work, we further investigate the issue by combining both preparation approaches in an attempt to increase the mass density of the core without losing the grain connectivity. Both factors are known to be the keys for high current-carrying ability. By using an addition of Mg+2B powder mixture to pre-reacted MgB$_2$ powder we introduce the required liquid ``welding" and ``filling" phase in between MgB$_2$ particles. We also show that the properties and amount of grain boundaries are the factors influencing vortex pinning and current-carrying ability in the pure MgB$_2$ material. In this work, we also discuss the effect of thermo-magnetic instability which turns out to be highly pronounced in the in-situ prepared superconductor.

\section{Experimental details}

Commercially available magnesium powder (350 mesh and 99+\% pure) and amorphous boron powder (99\% pure) with a composition of Mg+2B were mixed. The Mg+2B mixture was combined with commercial MgB$_2$ powder (Alfa-Aesar, containing about 2\% of secondary phases) in the following proportion (MgB$_2$)$_{1-x}$:(Mg+2B)$_x$, where $x$ is 0, 0.3, 0.5, 0.7, and 1.0. The particle size for Mg and MgB$_2$ powders was about the same $\sim 45\,\mu$m, whereas the B powder had a particle size of about 40~nm. The final mixtures were then ground in a mortar by hand for half an hour. The wires were prepared by the powder-in-tube technique: the mixtures were packed into iron tubes having outer and inner diameters of 7.6~mm and 5.1~mm, respectively. Both ends of the tubes were sealed by Fe-rod plugs. The density of the powder compression after the mechanical deformation (before heat treatment) was estimated to decrease with increasing $x$ from $1.81 \times 10^3$ kg/m$^3$ ($x = 0$) to $1.24 \times 10^3$~kg/m$^3$ ($x = 1$) which is simply due to the decreasing amount of the reacted powder having a higher mass density. The filled tubes were groove rolled into square wires with a square side of 3.3~mm. Approximately, 2 cm long pieces were cut from the wires for each proportion of the mixture and then sintered at 950$^{\circ}$C for 3 hours in a furnace with a continuous argon flow. The sintering was followed by a quench-cooling in liquid nitrogen. The iron sheath was then mechanically removed from all the samples to enable comparative magnetic measurements of the superconducting cores. The influence of the iron sheath shielding upon magnetic measurements is considered elsewhere \cite{paniron}. The samples were then polished to rectangular shapes having similar dimensions. The magnetization measurements were carried out in a MPMS SQUID magnetometer in a temperature range of 5-40~K within the applied magnetic field $|B_a| \le 5$~T. The field was applied parallel to the longest dimension of the sample, i.e., along the wire axis. A magnetic $J_c$ was derived from the half-width of the magnetization loops $\Delta M = (|M^+| + |M^-|)/2$ (with $M^+$ and $M^-$ being descending and ascending branches of the magnetization loop, respectively), using the following critical state model formula: $J_c = k \Delta M/d$, where $k = 12w/(3w-d)$ is a geometrical factor, with $d$ and $w$ being sample thickness and width, respectively. X-ray diffraction (XRD) analysis was used for control of the phase composition. Scanning electronic microscopy (SEM) was used for analysis of the core microstructure. The mass density of MgB$_2$ cores was estimated from data obtained by measurements of their weight and volume.

\section{Results and discussion}

\subsection{Microstructure and density} \label{ms}

\begin{figure*}
\includegraphics[scale=0.85]{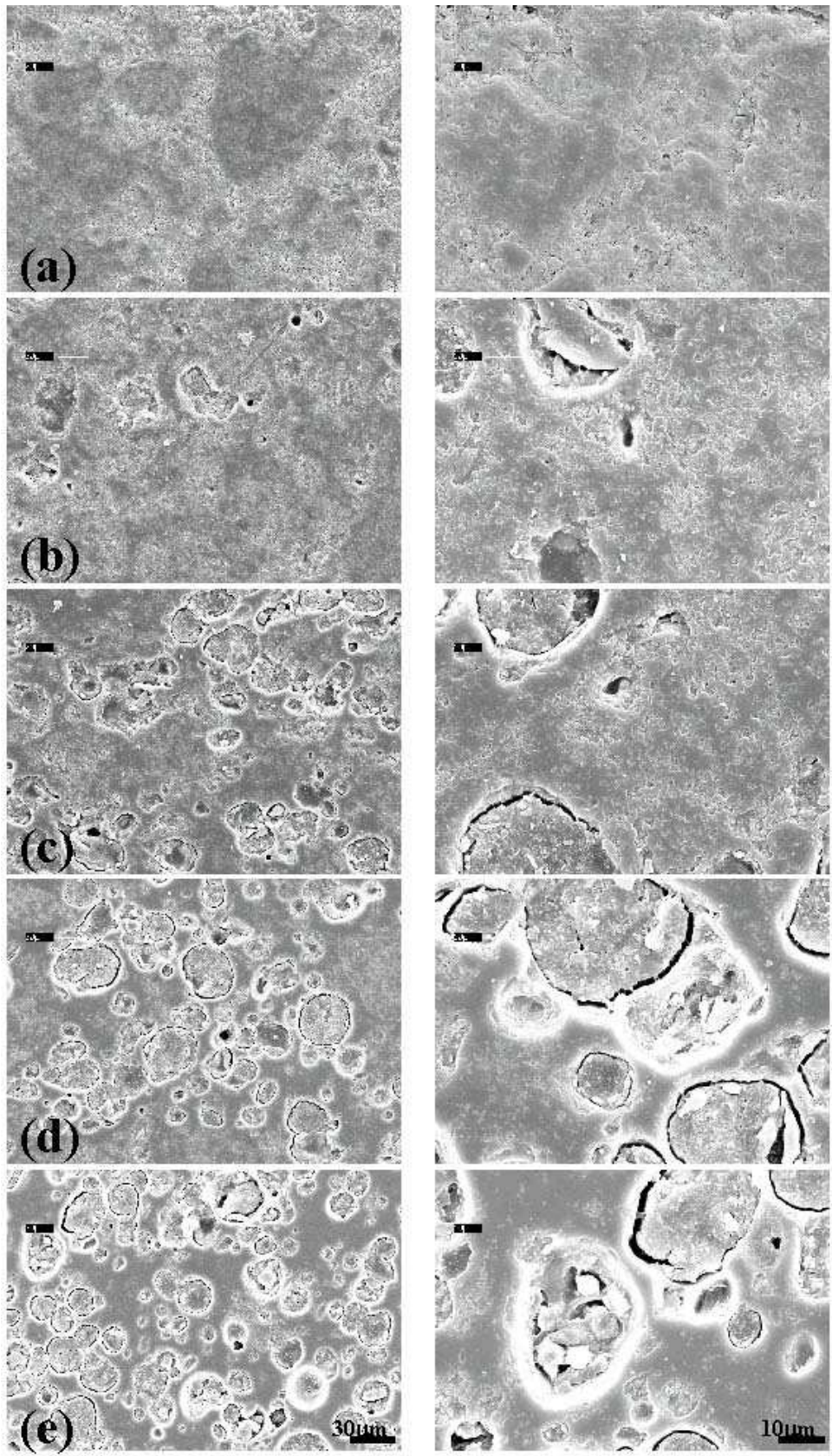}
\caption{\label{sem}Microstructure images of all the investigated samples with $x = 0$(a), 0.3(b), 0.5(c), 0.7(d), and 1(e). The left and right columns show lower and higher magnifications for each sample, respectively.}
\end{figure*}

Figure~\ref{sem} shows a series of lower (left) and higher (right) magnification SEM images taken for the samples studied. One can readily see the continuous micro-structural development of the core as $x$ increases from 0 to 1. For the sample with $x = 0$ the microstructure appears to be a relatively homogeneous formation, consisting of small particles (grains) $< 5\,\mu$m. Basically, the microstructure looks like well-compressed powder, rather than a granular material. The particle size appears to be much smaller than the initial particle size of the MgB$_2$ powder used for filling the iron tubes. No large (micro) scale voids are to be seen, however, a nano-scale porosity ($\le 1-2 \, \mu$m) is obvious for this sample. The porosity scale can be further reduced if either pressing or a higher overall deformation rate is used for mechanical processing. However, we note that at higher deformation rates iron becomes quite hard and brittle \cite{pan1}, which is not desirable for long wire manufacturing and applications. Iron annealing which could soften the material, is not suitable either, due to the rather high annealing temperature required, which is in the range of MgB$_2$ phase formation and can introduce secondary phases into the superconductor.

\begin{figure}
\begin{center}
\includegraphics[scale=0.34]{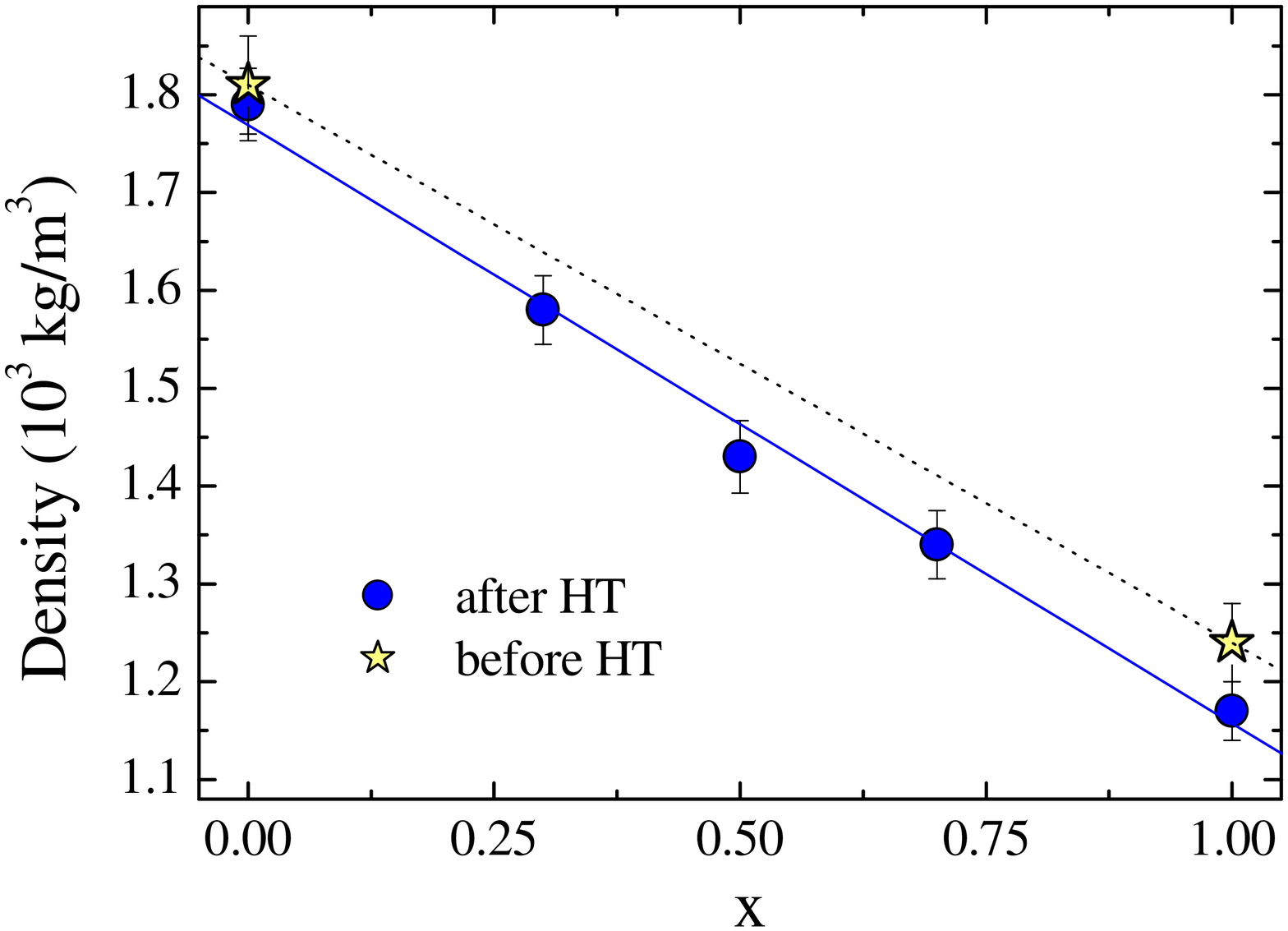}
\caption{\label{dens} Core density before and after heat treatment (HT). Lines are linear fits to the data, see also text.}
\end{center}
\end{figure}

After addition of the unreacted Mg+2B mixture (figure~\ref{sem}(b)), the core microstructure for the sample with $x = 0.3$ shows a trend towards becoming a granular material, rather than just being well-connected powder particles. Although the particle size is about the same as in the $x = 0$ case, the grains look better connected. Indeed, the liquid phase of the Mg+2B mixture during the sintering eventually turns into a MgB$_2$ matrix in which initial particles of pre-reacted MgB$_2$ powder are embedded. Correspondingly, a smaller amount of the nano-scale porosity is observable. Instead, a small volume of large scale voids (of the order of 10~$\mu$m) has appeared, as a result of a core shrinkage due to MgB$_2$ formation \cite{pan2}. The appearance of the voids did not allow us to achieve larger values of the overall mass density for the core (figures~\ref{dens}). A further increase in the amount of the Mg+2B mixture in the powder composition leads to a further reduction of the nano-scale porosity and a further development of the large scale voids (figures~\ref{sem}(c, d)). Any trace of the porosity disappears for the sample with $x = 0.7$. However, one can still see a fine granular structure of ex-situ particles within the in-situ formed MgB$_2$ matrix in between the voids. The matrix looks very dense in contrast to the samples with smaller $x$ values, but the {\it overall} density, due to the voids, turns out to be quite small in comparison with those samples. The sample with $x = 1$, in contrast to all the other samples, shows a monolithic core matrix around large scale voids (figure~\ref{sem}(e)). No small particles are observed within the matrix. Generally, the grain size of the in-situ formed MgB$_2$ superconductor is likely to be up to two orders of magnitude larger \cite{dou} than the ex-situ particles found in samples with $x < 1$. The volume occupied by the voids is the largest among all the samples investigated. Note also that this sample has the lowest overall mass density $\simeq 1.17 \times 10^3$~kg/m$^3$ (figure~\ref{dens}).

The estimated density of the samples after sintering as a function of the amount of the Mg+2B mixture added is shown in Fig.~\ref{dens}. As mentioned above, we also estimated the density for two samples with $x = 0$ and 1 before sintering. These estimations were more difficult because the core is rather loose at this stage, hence, the experimental error is slightly larger. The behavior after sintering is linear within the error of the estimation. In spite of having only two points for the unsintered samples we anticipated a linear dependence for those samples, as well. Notably, there is an {\it insignificant} difference between the curves. This means that after MgB$_2$ formation the core shrinks and the space created due to the difference in theoretical densities between the Mg+2B mixture ($2.0 \times 10^3$~kg/m$^3$) and MgB$_2$ ($2.63 \times 10^3$~kg/m$^3$) {\it directly} transforms to voids which can take up to 24\% of the overall core volume for the sample with $x = 1$. This conclusion is also supported by SEM observations (the left column in Fig.~\ref{sem} gives a comparative  idea on the volume of the voids formed). As a matter of fact, an experimental 30\% difference in the packing density between MgB$_2$ and Mg+2B mixture before HT in Fig.~\ref{dens} is in a good agreement with the theoretical density difference of 24\% mentioned above.

\begin{figure}
\begin{center}
\includegraphics[scale=0.34]{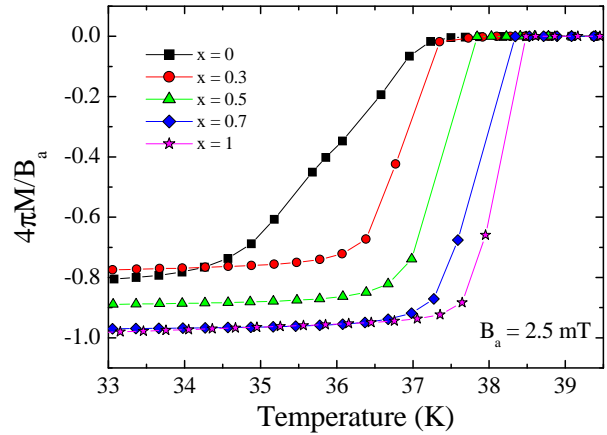}
\caption{\label{tc} Zero-field cooled state susceptibility measured on warming as a function of temperature. No demagnetizing factor was taken into account, but this would be expected to be very similar for our samples due to the similar dimensions. However, the differences in their microstructure explained in Subsection \ref{ms} may affect this expectation and explain the higher shielding factor (larger susceptibility $|4 \pi M/B_a|$) for the sample with $x = 0$ than for $x = 0.3$ at $T \le 34$~K.}
\end{center}
\end{figure}

XRD analysis basically showed the expected result: reflection peak patterns for all investigated samples are consistent with a well-formed  MgB$_2$ composition. A typical XRD pattern can be found, for example, in Fig.~1 in Ref.~\cite{pan1} for the Fe-sheathed tape. The same critical temperature onset $T_c^{\rm onset} = 38.6 \pm 0.1$~K defined at $4 \pi M/B_a = - 10^{-4}$ for all the samples further supports the structural similarity of both MgB$_2$ components: pre-reacted and formed in-situ. The difference in the onset of the transition temperature is negligible compared to the width $\Delta T_c$ of the transition. Indeed, the transition width significantly changes from $\Delta T_c \simeq 1.75$~K for $x = 1$ to $\Delta T_c \simeq 4.85$~K for $x = 0$ (Fig.~\ref{tc}), which is the first sign of the changing pinning properties described in the next subsection.
 
\subsection{Critical current density and vortex pinning}

\begin{figure}
\begin{center}
\includegraphics[scale=0.34]{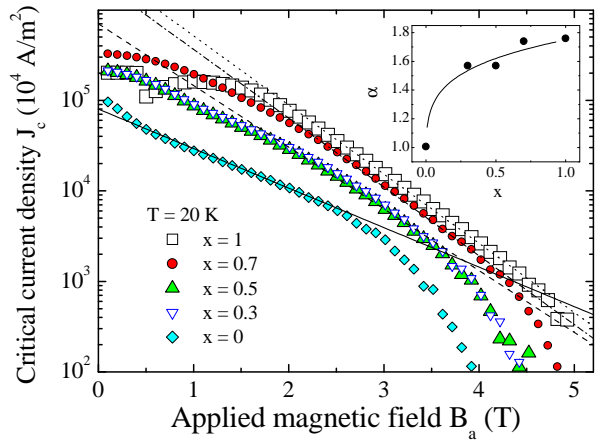}
\caption{\label{jc} The critical current density as a function of the applied magnetic field for all the investigated samples at $T = 20$~K. The lines show exponential fits to the linear intervals of the $J_c(B_a)$ dependences. The inset shows the pinning-dependent exponential pre-factor $\alpha$ versus $x$. The solid line in the inset is only a guide for the eye.}
\end{center}
\end{figure}

\begin{figure}
\begin{center}
\includegraphics[scale=0.34]{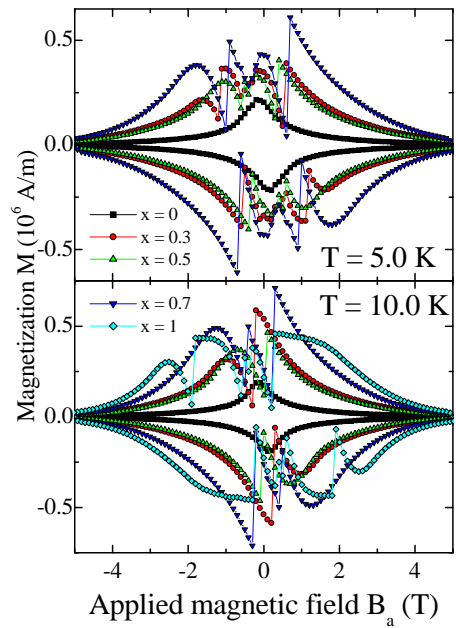}
\caption{\label{mag} Magnetization loops for the measured samples at $T = 5$~K and 10~K.}
\end{center}
\end{figure}

Figure~\ref{jc} shows the critical current density as a function of the applied magnetic field $B_a$ for all the investigated samples at $T = 20$~K. The smallest $J_c$ was measured for the sample with $x = 0$ (pure ex-situ preparation) over the entire field and temperature range investigated. The general trend is straightforward: the larger the $x$, the higher the $J_c$. At $B_a < 1.2$~T, we can anticipate a nominally higher $J_c$ for the $x = 1$ (pure in-situ preparation) sample than for the other samples, although it is masked by the thermo-magnetic flux-jump instabilities. The flux-jumps at this temperature are absent for all the other samples prepared, most probably indicating that the pure in-situ preparation has the largest shielding current across well-connected, presumably large MgB$_2$ grains amongst all the samples. This is also in agreement with the fact that the highest Meissner shielding factor (the largest susceptibility $|4 \pi M/B_a|$) was observed below the transition temperature for this sample (figure~\ref{tc}). At lower temperatures, flux-jumps appear for all the samples ($0.3 \ge x \ge 1$) except the one with $x = 0$ (figure~\ref{mag}). As can be clearly seen from the magnetization curves, the flux-jumps do not appear for this sample at all, down to at least $T = 5$~K. This fact, along with the very low $J_c(B_a)$, is likely to indicate the small ex-situ grain size and a lack of connectivity (transparency) between these grains. A significant increase in $J_c$ is achieved by an addition of the Mg+2B powder and the consequent in-situ reaction assistance. However, already starting from $x = 0.3$, $J_c(B_a)$ curves as a function of $x$ become less sensitive to the level of the Mg+2B powder addition (figure~\ref{jc}).

The irreversibility field $B_{\rm irr}$ defined at $J_c = 100$~A/cm$^2$ increases with increasing $x$ (figure~\ref{jc}). For the sample with $x = 0$, $B_{\rm irr} = 3.9$~T, whereas for $x = 1$, $B_{\rm irr} \simeq 5.2$~T. Furthermore, one can see that for the pure ex-situ prepared sample the $J_c(B_a)$ curve has a rapid drop at about 2.5~T defined at the point where an exponential fit to the linear $J_c(B_a)$ interval in the $\log J_c$-$B_a$ scale diverges from the $J_c(B_a)$ measured (figure~\ref{jc}). The drop shifts to higher fields with increasing $x$. For the sample with $x = 1$ there is only a slight indication of a drop at about $B_a = 4.6$~T which is quite close to the irreversibility (depinning) field. In the samples with intermediate levels of $x = 0.3$ and 0.5 the drop occurs at about 3.6~T. The drop is likely to occur when the ``weak links" stop contributing to the overall current of the superconductor as suggested in \cite{dou}. These ``weak links", even in the purely ex-situ prepared wire, appear to be rather strong if they are compared to high-$T_c$ BSCCO tapes, where a similar downturn is usually expected to occur below 0.1~T \cite{joseph}. However, we should note that the approach used in \cite{dou} to calculate $J_c$ for intra-granular currents at high fields did not work in our case for the sample with $x = 0$, where the average particle size ($\sim 3 \, \mu$m) could be estimated with the help of SEM image analysis (figure~\ref{sem}(a)). According to our $J_c$ calculation for the sample with $x = 0$, the critical current density at high fields appeared to be higher than the value of $J_c(0)$. An inhomogeneous particle compression inside the core, forming some larger and better connected conglomerates, might be responsible for the failure of the intra-granular current determination. Note that in Ref.~\cite{dou}, where an in-situ preparation samples were discussed, a grain size of $170 \times 260 \, \mu$m was used for the calculation.

An additional bit of information can provide the field dependence of $J_c$. To a large extent, $J_c(B_a)$ can be described by an exponential function ($J_c \sim \exp(-\alpha B_a$), where the pre-factor $\alpha$ is a pinning dependent parameter, changing with $B_a$, $T$, and the microstructure). The lines shown in figure~\ref{jc} are exponential fits to the corresponding linear $J_c(B_a)$ intervals in the $\log J_c$-$B_a$ scale, which means that in this field range $\alpha$ depends {\it only} on the core microstructure. Correspondingly, in the inset to Fig.~\ref{jc} the pre-factor $\alpha$ as a function of $x$ is shown. One sees that the sample with $x = 0$ has the slowest $J_c$ degradation within the fitted interval just before the $J_c$ drop. We argue that this is due to the pinning of vortices along grain boundaries in the samples. The more grain boundaries, the slower the pinning decrease until the weak links stop contributing to the total current, which happens at rather high fields.

Thus, on the basis of microstructure analysis and magnetic measurements we can propose the following scenario for describing the $J_c$ behavior as a function of $x$. The smaller $x$ is, the more small ex-situ grains and, correspondingly, the more boundaries there are in the core. These boundaries possess rather low current-carrying transparency, since the level of the overall current transparency decreases with decreasing $x$. It is evident from the smaller values of the $J_c$ measured for smaller values of $x$ (with the lowest $J_c$ obtained for the $x = 0$ sample). However, a certain level of connectivity between the grains survives up to rather high field values (until the $J_c$ drop). A large amount of grain boundaries ensures the largest amount of pinning centers in the sample with $x = 0$. Hence, the slowest $J_c$ degradation with field is measured. As soon as $B_a$ starts breaking ``weak links" at the grain boundaries, the vortices start rearranging themselves so that a significant $J_c$ drop occurs. In addition, to complete the description of the sample with $x = 0$ (purely ex-situ prepared sample) we explain the first $J_c$ decrease at $B_a < 0.3$~T which is not measured for any other sample investigated in this work. SEM investigation shows that the core has both loosely and densely packed particle conglomerates (observable in figure~\ref{sem}(a)), which leads to the failure of the grain size related $J_c$ estimation at high fields. These facts suggest that the connectivity between grains is quite inhomogeneous. Therefore, the first $J_c$ drop below 0.3~T is likely to be due to the loosely packed (porous) regions with presumably the weakest grain connectivity, whereas the second drop at higher fields is due to decoupling of the stronger packed and larger volume regions.

As $x$ is increased, an enhanced connectivity (transparency) between the small ex-situ grains is introduced through formation of the in-situ matrix consisting of much larger grains. On the other hand, the amount of the {\it pinning} boundaries is reduced. As a consequence of the interplay between transparency and amount of pinning boundaries, one achieves larger $J_c$ values with their faster degradation as a function of $B_a$ (see the inset in figure~\ref{jc}). For the sample with $x = 1$, the effect of the boundaries is minimal: maximum boundary transparency and minimal quantity of boundaries are responsible for respectively the highest $J_c$ and for the fastest $J_c(B_a)$ degradation. The maximum boundary transparency and the highest $J_c$ are also responsible for the strongest thermo-magnetic instabilities observed in the widest field range in this purely in-situ prepared sample (figure~\ref{mag} for $T = 10$~K). It is worth noting that the large scale voids are expected to affect neither actual $J_c$ value, nor the flux-jumps. The voids do, however, affect the overall $I_c$ of the wires by reducing the superconducting volume and the effective super-current path cross section. This means that the real $J_c$ in the samples with $x > 0$ can be up to 24\% higher for the sample with $x = 1$, compared to the calculated values shown in figure~\ref{jc}.
 
Undoubtedly, larger values of $B_{\rm irr}$ and $J_c$ in the entire applied field range illustrate a significant advantage of the pure in-situ and of in-situ assisted wire preparation over the pure ex-situ procedure. However, the danger of the flux-jump effect becomes stronger with increasing $x$ (Figs.~\ref{jc} and \ref{mag}). As can be seen, at  $T < 20$~K, the flux-jumps become apparent not only for the sample with $x = 1$, but also for the other samples, except for $x = 0$. Therefore, these instabilities provide an example of how nominally better samples can actually be worse for practical applications. Indeed, some applications of wires which might have an unstable superconducting state would be highly problematic. Moreover, our preliminary results \cite{paniron2} show that the flux-jump effect is even stronger in the iron sheath, occurring over a wider field range, than in the case presented in this work, where the sheath is removed from the wires. This is quite understandable, since the iron has low thermal conductivity, which worsens the thermal sink environment of the superconductor, compared to the environment of the steady cold helium gas flow around the bare core. Therefore, the chance of experiencing instabilities in such an environment is higher \cite{inst}. Small resistivity sheath materials, such as copper, silver, and aluminium, which could stabilize the wires \cite{stab}, were shown to be too soft to form an acceptable level of $I_c$ and chemically incompatible for in-situ MgB$_2$ core preparation \cite{pan1}. Furthermore, a highly promising drastic enhancement of $J_c(B_a)$ performance by SiC nano-particle doping in MgB$_2$ wires has been obtained by the in-situ preparation method, as was recently reported by authors \cite{pansic1,pansic2}. It is at present unclear whether it would be possible to obtain similar characteristics for MgB$_{2-x}$(SiC)$_x$ superconducting wires prepared by the ex-situ procedure. On the basis of the present study, we can anticipate that the pinning effect of SiC nano-doping might be overruled by the grain boundary effect described above. However, further study is needed to clarify the issue. Generally, the literature shows that the flux-jumps are usually obtained for in-situ prepared samples \cite{pan3,xiaolin,pansic1,pansic2,dou}, whereas for ex-situ preparation the flux-jumps are rarely observed. Moreover, the flux-jumps disappear if any second stage mechanical treatment that crashes the core is involved \cite{pan3}. It is further shown in Ref.~\cite{pan3} that a consecutive heat-treatment is not able to re-join the grains, so that the chance for flux-jump occurrence in those samples is low.

A lower current transparency of the grain boundaries in the case of the ex-situ preparation than in the case of the in-situ preparation can also explain the usually larger resistivity, measured just above the critical temperature, for the ex-situ prepared samples (typically $> 50 \, \mu \Omega$cm) than for the in-situ preparation (0.4-1~$\mu \Omega$cm), see \cite{16} and corresponding references therein. Assuming this is the case for our samples and taking into account the fact that the irreversibility field $B_{\rm irr}$ increases with increasing $x$, we conclude that the difference in the resistance is due to different grain boundary transparencies obtained by the in-situ or ex-situ procedure. Therefore, we expect that these different preparation procedures would produce superconductors belonging either to the clean- or dirty-limit. A procedure for ``polluting" clean-limit MgB$_2$ samples by annealing them in Mg vapor in order to obtain a higher $B_{\rm irr}$, can be found in \cite{16}.

\section{Conclusion}

From studying samples with different initial powder composition (MgB$_2$)$_{1-x}$:(Mg+2B)$_x$ where $x$ varies from 0 to 1, we found that the critical current density as a function of the applied field is larger over the entire field range for samples with $x > 0$. An addition of the unreacted Mg+2B mixture to the MgB$_2$ powder enhances the transparency of the grain boundaries, introducing an MgB$_2$ matrix that is formed in-situ with embedded small MgB$_2$ ``ex-situ" particles (for $x < 1$). The grain boundaries are found to be responsible for the pinning of vortices. The larger $x$ is, the smaller the amount of {\it pinning} boundaries, so that $J_c(B_a)$ decays faster. A pure in-situ reaction ($x = 1$) forms a core with nominally the best characteristics, despite the fastest field degradation. However, we emphasize the following features which may affect the employment of the pure in-situ procedure for commercial wire fabrication. (i) Thermo-magnetic instabilities which are highly undesirable for high-current and field applications are strongly pronounced within quite a wide field range up to $T = 20$~K which is considered to be a benchmark for this material. (ii) The other feature is large scale void formation in the core which can lead up to 24\% lower overall critical current than the theoretically achievable one.

\begin{acknowledgments}
We would like to thank T. Silver for careful reading of the manuscript and valuable critical remarks. This work was financially supported by the Australian Research Council.
\end{acknowledgments}


\begin{thebibliography}{99}

\bibitem{grasso}Grasso G, Malagoli A, Ferdeghini C, Roncallo S, Braccini V, Cimberle M R, Siri A S 2001 Appl. Phys. Lett. {\bf 76} 230.
\bibitem{fujii}Fujii H, Kumakura H and Togano K 2001 Physica C {\bf 363} 237.
\bibitem{suo}Suo H L, Beneduce C, Dhall\'e M, Musolino N, Genoud J-Y and Fl\"ukiger R 2001 Appl. Phys. Lett. {\bf 79} 3116.
\bibitem{pan2}Zhou S, Pan A V, Liu H and Dou S 2002 Physica C {\bf 382} 349.
\bibitem{pan3}Zhou S, Pan A V, Liu H, Horvat J and Dou S 2002 Supercond. Sci. Technol. {\bf 15} 1490.
\bibitem{xiaolin}Wang X L, Soltanian S, Horvat J, Liu A H, Qin M J, Liu H K, and Dou S X 2001 Physica C {\bf 361} 149.
\bibitem{martinez}Martinez E, Angurel L A and Navarro R 2002 Supercond. Sci. Technol. {\bf 15} 1043.
\bibitem{pansic1}Dou S X, Pan A V, Zhou S, Ionescu M, Liu H K and Munroe P R 2002 Supercond. Sci. Technol. {\bf 15} 1587.
\bibitem{pansic2}Dou S X, Pan A V, Zhou S, Ionescu M, Wang X L, Horvat J, Liu H K and Munroe P R, ``Superconductivity, critical current density, and flux pinning in MgB$_{2-x}$(SiC)$_{x/2}$ superconductor after SiC nano-particle doping", Preprint Cond-mat/0207093.
\bibitem{paniron}Pan A V, Zhou S, Dou S and Liu H, ``Direct visualization of iron sheath shielding effect in MgB$_2$ superconducting wires", Preprint Cond-mat/0205253.
\bibitem{pan1}Zhou S, Pan A V, Ionescu M, Liu H and Dou S 2002 Supercond. Sci. Technol. {\bf 15}, 236 (2002).
\bibitem{dou}Dou S X, Wang X L, Horvat J, Milliken D, Li A H, Konstantinov K, Collings E W, Sumption M D and Liu H K 2001 Physica C {\bf 361} 79.
\bibitem{joseph}Horvat J, Dou S X, Liu H K and Bhasale R 1996 Physica C {\bf 271} 51.
\bibitem{paniron2}Pan A V et al., unpublished.
\bibitem{inst}Mints R G and Rakhmanov A L 1977 Usp. Fiz. Nauk (Rus.) {\bf 121} 499; 1981 Rev. Mod. Phys. {\bf 53} 551.
\bibitem{stab}Kontorowitz A R and Stekly Z J 1965, Appl. Phys. Lett. {\bf 6} 56.
\bibitem{16}Braccini V, Cooley L D, Patnaik S, Manfrinetti P, Palenzona A, Siri A S, Larbalestier D C 2002 Appl. Phys. Lett. {\bf 81} 4577.

\end{thebibliography}
\end{document}